\documentclass[conference]{IEEEtran}
\ifCLASSINFOpdf
  \usepackage[pdftex]{graphicx}
\else
\fi
\begin{document}
%
\title{IMP Science Gateway: from the Portal to the Hub of Virtual Experimental Labs in Materials Science}

\author{\IEEEauthorblockN{Yuri Gordienko*, Lev Bekenev, Olexandra Baskova,\\Olexander Gatsenko, Elena Zasimchuk}
\IEEEauthorblockA{G.V.Kurdyumov Institute for Metal Physics,\\National Academy of Sciences\\
Kyiv, Ukraine\\
Email: *gord@imp.kiev.ua}
\and
\IEEEauthorblockN{Sergii Stirenko}
\IEEEauthorblockA{\\High Performance Computing Center,\\
National Technical University of Ukraine "KPI",\\
Kyiv, Ukraine\\
Email: sergii.stirenko@hpcc.kpi.ua}}


%


\maketitle

\begin{abstract}
The user communities outside of computer science itself (for example, in materials science, physics, chemistry, biology, nanotechnologies, etc.) would like to access different software components and various distributed computing infrastructures (DCIs) (like grids, clouds, clusters) in a such transparent way, that they do not need to learn their peculiarities. In this context, "science gateway" (SG) ideology means a user-friendly intuitive interface between scientists (or scientific communities) and different software components + various DCIs, where researchers can focus on their scientific goals and less on peculiarities of software/DCI. "IMP Science Gateway Portal" (http://scigate.imp.kiev.ua) for complex workflow management and integration of distributed computing resources (like clusters, service grids, desktop grids, clouds) is presented. It is created on the basis of WS-PGRADE and gUSE technologies, where WS-PGRADE is designed for science workflow operation and gUSE - for smooth integration of available resources for parallel and distributed computing in various heterogeneous distributed computing infrastructures (DCI). The typical scientific workflows with possible scenarios of its preparation and usage are presented. Several typical use cases for these science applications (scientific workflows) are considered for molecular dynamics (MD) simulations of complex behavior of various nanostructures (nanoindentation of graphene layers, defect system relaxation in metal nanocrystals, thermal stability of boron nitride nanotubes, etc.). The user experience is analyzed in the context of its practical applications for MD simulations in materials science, physics and nanotechnologies with available heterogeneous DCIs. The most crucial advantages are as follows: powerful and flexible workflow construction tools, which allow users to apply LEGO-style for creation of complex tunable chains with de facto standard components (like R, LAMMPS, AtomEye, ffmpeg) and customized components (like debyer, Pizza, etc.); smooth access to heterogeneous DCI and software; division of user roles (administrators: operators of portals; power users: principal scientists; end users: scientists, students); user friendly and intuitively understandable GUI (with additional modules, ad hoc changes, etc.); extensive manuals and permanently working fora and helpdesks; short learning curve for usual scientists without extensive knowledge in computer science. Some non-critical drawbacks of the approach include: scarce "stdout"and "stderr" information, tacit context help for beginners, some unavoidable upgrade issues, etc. In conclusion, the "science gateway" approach - workflow manager (like WS-PGRADE) + DCI resources manager (like gUSE)- gives opportunity to use the SG portal (like "IMP Science Gateway Portal") in a very promising way, namely, as a hub of various virtual experimental labs (different software components + various requirements to resources) in the context of its practical MD applications in materials science, physics, chemistry, biology, and nanotechnologies.
\end{abstract}


%
\IEEEpeerreviewmaketitle

\section{Introduction}
Recently research boom in nanotechnologies has started to search and investigate new nanoscale materials, like nanocrystals, graphene, carbon nanotubes (CNT), boron nitride nanotubes (BNNT), and their complexes, because of their unique properties. Molecular dynamics (MD) simulations of nanoscale structures, processes, and properties are very promising in the wide range of physical parameters, because of necessity to investigate wide range of parameters (so-called ``parameter sweeping''). MD simulations with such ``parameter sweeping'' can be carried out in the available heterogeneous distributed computing infrastructure (DCI) with desktop grids, service grids, clusters, supercomputers, cloud resources. The recent advances in computing algorithms and infrastructures, especially in development of DCIs, allow us to solve these tasks efficiently without expensive scaling-up, especially by means of the science gateway (SG) technology on the basis of WS-PGRADE platform for workflow management and gUSE technology for integration of DCI-resources \cite{wspgrade2012}. Here we demonstrate the capabilities of these technologies realized for MD simulations and further data post-processing on the basis of LAMMPS package for MD simulation \cite{lammps1995}, and other packages like R for statistical analysis \cite{R2008}, Pizza.py Toolkit for manipulations with atomic coordinates files \cite{pizza2005}, AtomEye for visualisation of atomic configurations \cite{AtomEye2003}, debyer for simulation of X-ray diffraction (XRD) and neutron diffraction (ND) analysis (https://code.google.com/p/debyer/), etc. Several typical workflows were created for simulation of several physical processes with various demands for the computing resources: tension of metal nanocrystals under different physical conditions, tension of ensemble of metal nanocrystals under the same conditions, manipulations with complex nanostructures like indentation of graphene membranes, thermal stability of BNNT, etc.

\section{Background and Related Works}
Scientists from user communities outside of computer science itself (for example, in materials science, physics, chemistry, biology, nanotechnologies, etc.) would like to access various DCIs (grids, clouds, clusters) and concentrate on their scientific applications themselves in a such transparent way, that they do not need to learn the peculiarities of these DCIs. In general, ``science gateway'' (SG) implies a user-friendly intuitive interface between scientists (or scientific communities), various software (standard and/or proprietary) and heterogenous DCIs. Usually, SGs are not specialized for a certain scientific area and scientists from many different fields can use them. By means of SGs researchers can focus on their scientific goals and less on assembling the required DCI components. The most important aspects of SG are: a simplified intuitive graphical user interface (GUI) that is highly tailored to the needs of the given scientific community; a smooth access to national and international computing and storage resources; collaborative tools for sharing scientific data on national and international scale. Several SG technologies and frameworks are known (like ASKALON \cite{askalon2005}, MOTEUR \cite{moteur2008}, WS-PGRADE/gUSE/DCI-bridge \cite{wspgrade2012}), which use different enabling components and technologies: web application containers (Tomcat, Glassfish, etc.), portal or web application frameworks (Liferay, Spring, Drupal, etc.), database management systems (MySQL, etc.), workflow management systems \cite{wfreview:deelman2009}. Here usage of WS-PGRADE/gUSE/DCI-bridge is described with some examples in physics, materials science, nanotechnologies. The architecture of science gateway ideology on the basis of gUSE+WS-PGRADE+DCI-Bridge technologies (https://guse.sztaki.hu) is described in details elsewhere \cite{wspgrade2012}. In short, WS-PGRADE portal technology is a web based front end of the gUSE infrastructure. gUSE (grid and cloud User Support Environment) is a permanently improving open source SG framework developed by Laboratory of Parallel and Distributed Systems (MTA-SZTAKI, Budapest, Hungary), that provides the convenient and easy access to DCI. It gives the generic purpose, workflow-oriented GUI to create and manage workflows on various DCIs. WS-PGRADE supports development and submission of distributed applications executed on the computational resources of DCIs. The resources of DCIs (including clusters, service grids, desktop grids, clouds) are connected to the gUSE by a single point back end, the DCI-Bridge \cite{wspgrade2012}.

\section{IMP Science Gateway Internals}
\begin{figure}[hbtp]
  \begin{center}
    \includegraphics[width=0.5\textwidth]{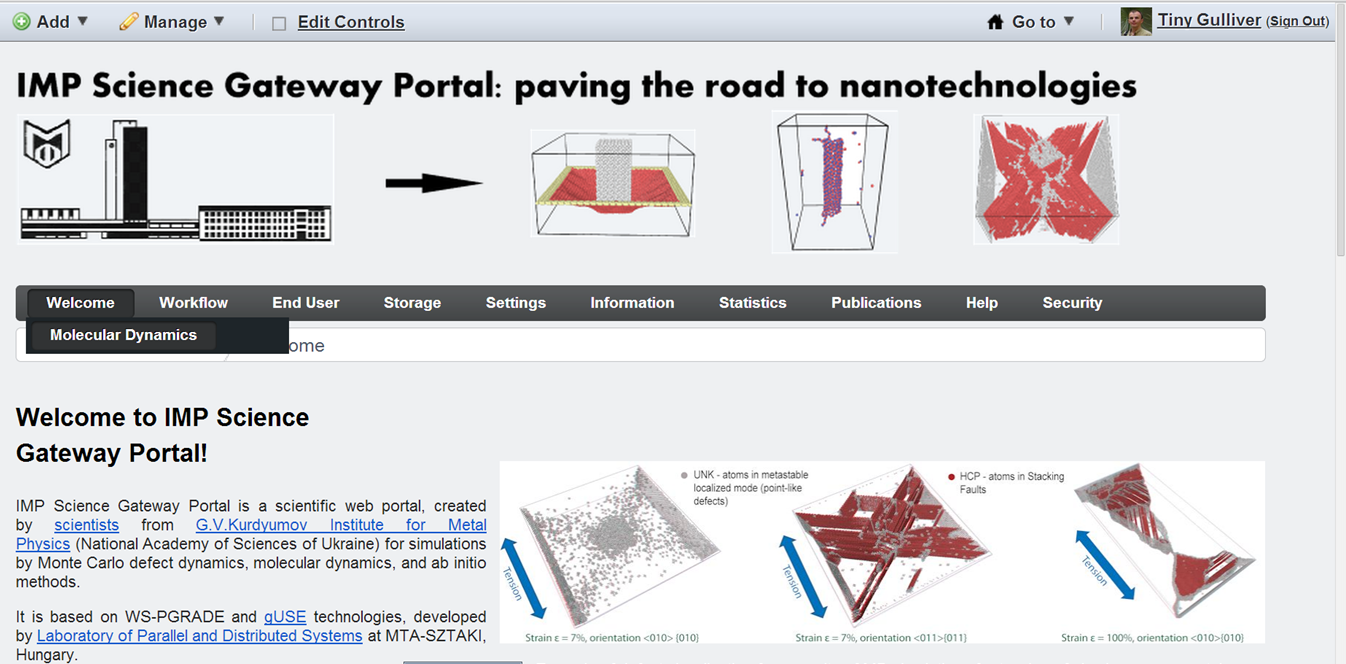}
    \caption{The graphical user interface of ``IMP Science Gateway Portal''.}
    \label{fig:Portal_GUI}
  \end{center}
\end{figure}

``IMP Science Gateway Portal'' \cite{GordienkoHPCUA2013} (Fig. \ref{fig:Portal_GUI}) is based on the WS-PGRADE/gUSE/DC-bridge framework \cite{wspgrade2012}, which is the complex system with many features. It distinguishes different user communities: administrators, power-users and end-users. The administrators operate portal itself. The power-users can develop new workflow applications by means of various workflow features (graph, abstract workflow, template, application). The workflows and their components are stored in built-in repository and published by the application developers, which can be shared among end-users. It provides scientific workflow management for both application developers and end users, and supports various DCIs including clusters, service grids, desktop grids and cloud.

\subsection{Workflow Management}
\begin{figure}[h!]
  \begin{center}
    \includegraphics[width=0.4\textwidth]{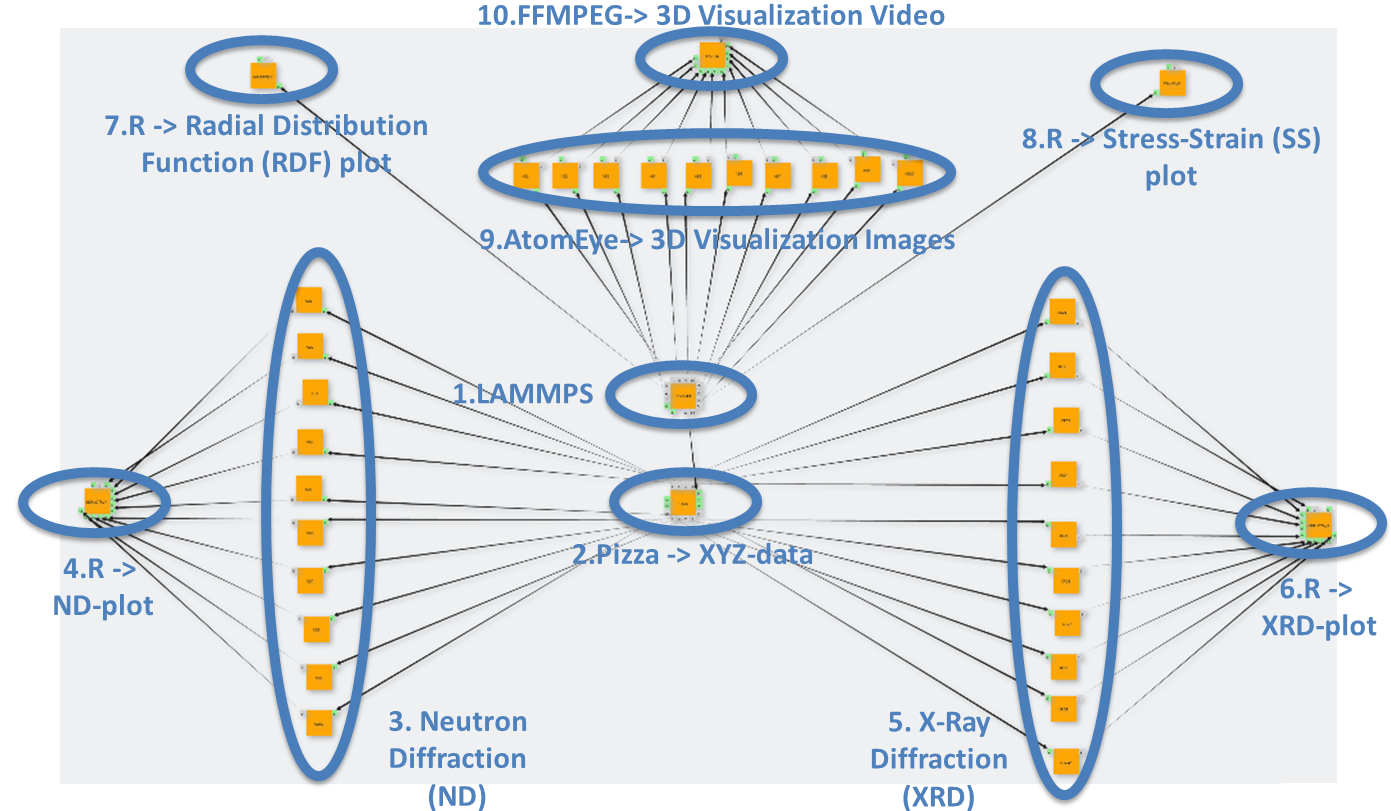} a)
    \includegraphics[width=0.4\textwidth]{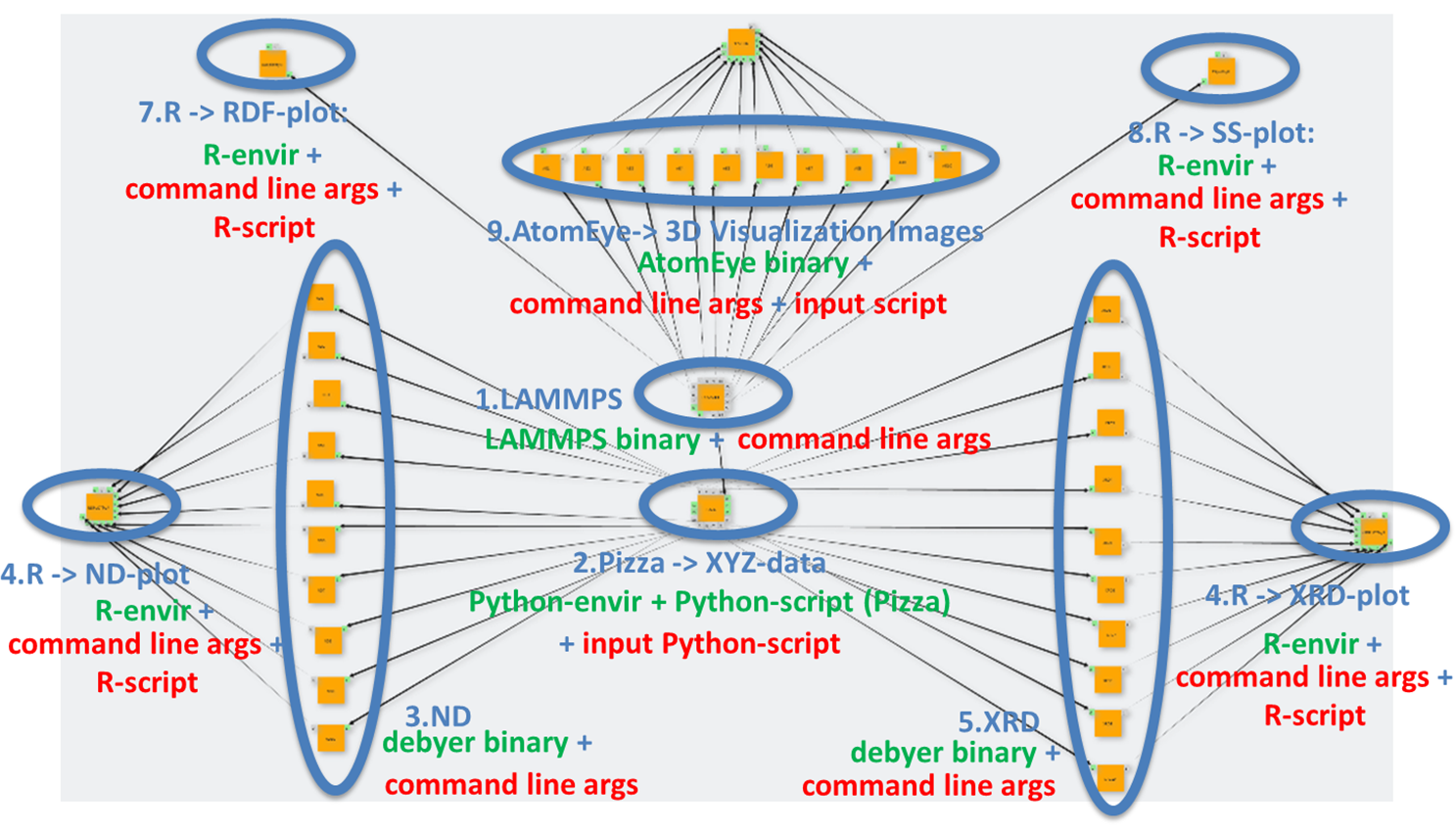} b)
    \includegraphics[width=0.4\textwidth]{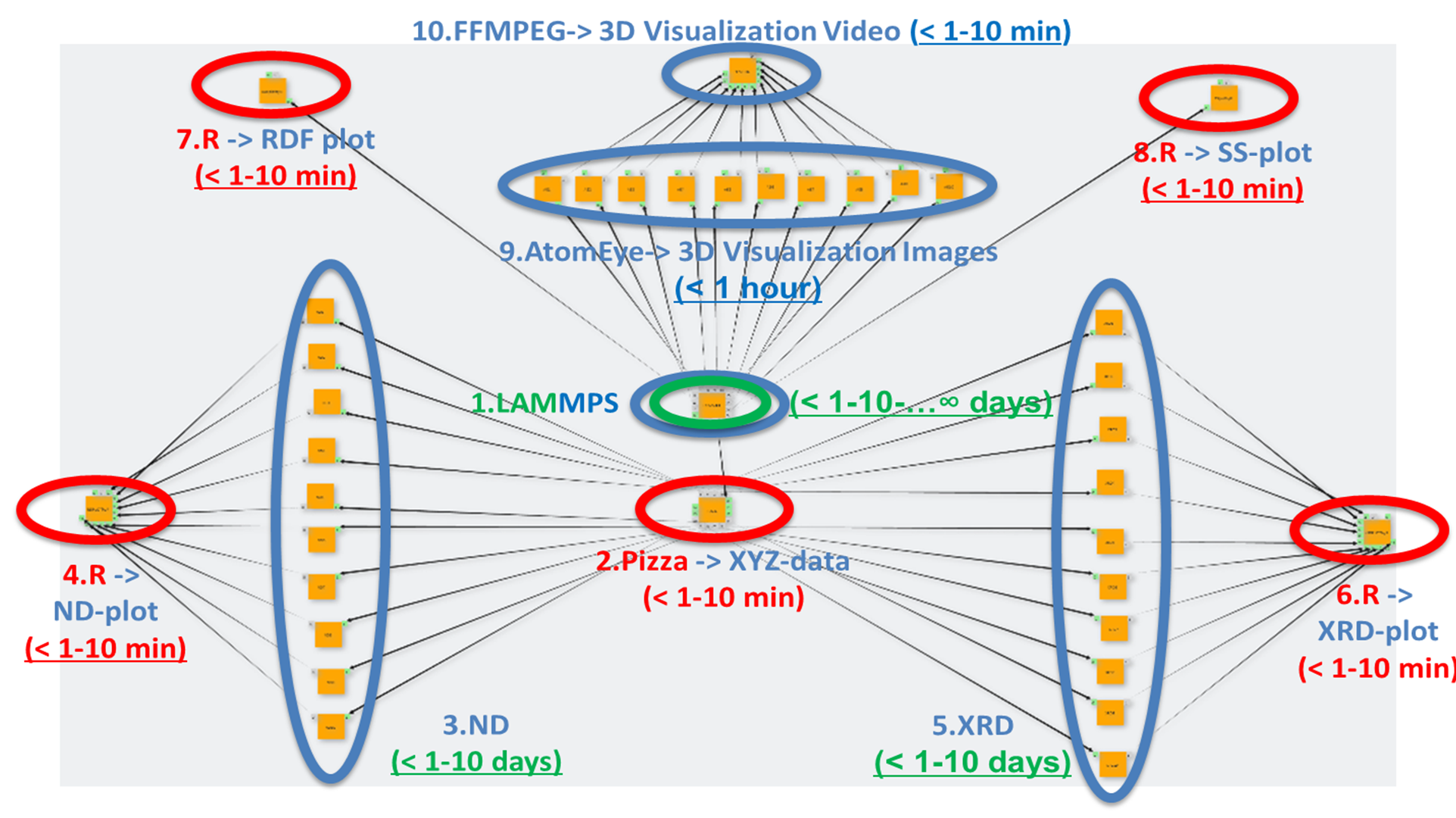} c)
    \includegraphics[width=0.4\textwidth]{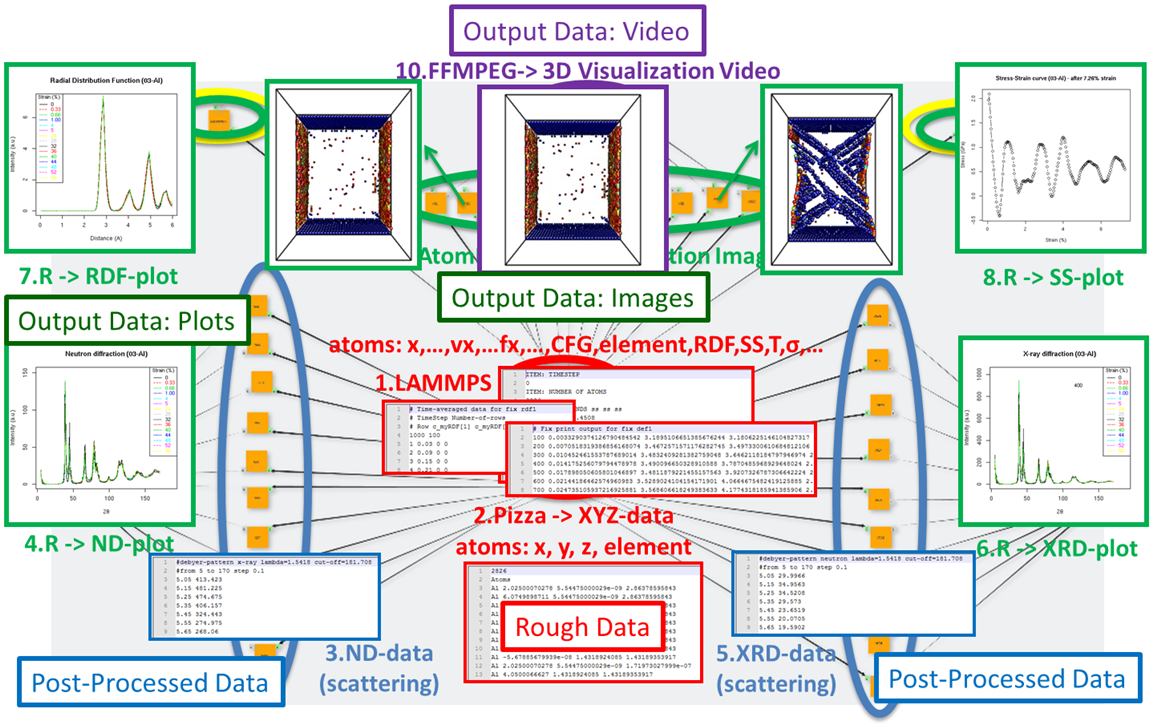} d)
    \caption{The scheme of workflow created by means of WS-PGRADE workflow management technology for MD simulations: (a) main components (see their description in the text); (b) invariant parts like executables, environmental parameters (green color) and variable parts like input data, scripts (red color); (c) job runtimes and resources needed: short runtimes on a server side (red color), medium --- on DCI (blue color), long --- on cluster (green color); (d) output data: rough text data (coordinates and velocities of atoms) of the huge size(red color), post-processed text data of the medium size (blue color), output image data (plots, 3D images) of the small size (green color), and output 3D video data of the small size (violet color).}
    \label{fig:AL_WF_scheme}
  \end{center}
\end{figure}

``IMP Science Gateway Portal'' leverages easy and simple GUI utilities for power-users (application developers) to create (easily in LEGO-style manner) the complex workflows (Fig.~\ref{fig:AL_WF_scheme}a) from separate simple components, which can be standard or proprietary executables, scripts in various languages, and command-line arguments (Fig.~\ref{fig:AL_WF_scheme}b). Application developers can design and create complex workflows with taking into account various runtimes and resources needed for each component (Fig.~\ref{fig:AL_WF_scheme}c), and output data type and size (Fig.~\ref{fig:AL_WF_scheme}d). End-users can submit, monitor, save, upload, and retrieve various workflows with one or many jobs, and obtain intermediate or final results (Fig.~\ref{fig:AL_WF_scheme}d).

The whole process of the complex workflow creation and management can be demonstrated by the MD simulation of tension of metal nanocrystals (Fig.~\ref{fig:AL_WF_scheme}a) with the following steps:
(1) to register/sign in ``IMP Science Gateway Portal''; (2) to create a graph scheme for the basic components of the workflow (yellow bricks with small input-output ports-squares in Fig.~\ref{fig:AL_WF_scheme}a); (3) to create a workflow from the graph; (4) to configure the workflow (links between yellow bricks in Fig.~\ref{fig:AL_WF_scheme}a); (5) to run the workflow; (6)
 to monitor the current status of execution. The complex workflow (Fig.~\ref{fig:AL_WF_scheme}) for this application contains several independent components (LAMMPS package, Pizza.py Toolkit, AtomEye package, debyer, R-package, FFMPEG), which can be combined in various sets for the purposes of the end users \cite{cgw2011,Gatsenko2013}. The end-users can check the intermediate results: after MD simulation by LAMMPS package (tables in the red rectangles in the center of Fig.~\ref{fig:AL_WF_scheme}d); converted by Pizza.py Toolkit script package for post-processing (table in the red rectangle in the bottom of Fig.~\ref{fig:AL_WF_scheme}d); visualizations of atomic positions obtained by AtomEye package (images in the top part of Fig.~\ref{fig:AL_WF_scheme}d); post-processed results of X-ray diffraction (XRD) and neutron diffraction (ND) analysis obtained by debyer (tables in the blue rectangles in the bottom part of Fig.~\ref{fig:AL_WF_scheme}d); final results after XRD and ND analysis, coordination number (CN) and radial distribution function (RDF) analysis, and stress-strain (SS) analysis obtained by R-package (plots in the green rectangles at the left and right sides of Fig.~\ref{fig:AL_WF_scheme}d); the final post-processed video of visualizations of atomic positions obtained by FFMPEG package (video in the violet rectangle in the top part of Fig.~\ref{fig:AL_WF_scheme}d).

\subsection{Integration of Computing Resources}
To run jobs gUSE allows to compose various resources and execute workflows on various DCIs: clusters (PBS, LSF), cluster grids (ARC, gLite, GT2, GT4, GT5, UNICORE), supercomputers (e.g. via UNICORE), desktop grids (BOINC) and clouds (via CloudBroker Platform and GAE). At the moment ``IMP Science Gateway Portal'' uses local resources (server side), several clusters (PBS) with different configuration and policies, tests desktop grids (BOINC) and cluster grids (ARC), and plans to use supercomputers and clouds. In this context, the different requirements for the various components were taken into account as to their possible location, runtime, output file size (see Table~\ref{table:Components}).

%
\begin{table}[!h]
\caption{The components of the general workflow (Fig.~\ref{fig:AL_WF_scheme})}
\label{table:Components}
\centering
\begin{tabular}{|c|c|c|c|}
\hline
\textbf{Component} & \textbf{Location} & \textbf{Runtime} & \textbf{Output Format (Size)} \\
\hline
LAMMPS & Cluster, DCI & 1-10 days & Text (1-10 GB)\\
\hline
R-package & Server (Cluster, DCI) & 1-10 min & Plot ($<1$MB)\\
\hline
Pizza-script & Server (Cluster, DCI) & 1-10 min & Text (1-10 GB)\\
\hline
AtomEye & Cluster, DCI & $<1$hour & 3D Image ($<1$MB)\\
\hline
FFMPEG & Cluster, DCI & 1-10 min & 3D Video ($<10$MB)\\
\hline
debyer & Cluster, DCI & 1-10 days & Text (10-100 MB)\\
\hline
\end{tabular}
\end{table}

\subsection{User Management}
The development and execution stages are separated between users with different background and responsibility (power-users and end-users) in the next two main user groups:
(1) ``end users'' --- the scientists without extensive knowledge of computer science with limited power and permissions (like common scientists in materials science, physics, chemistry, etc.), usually they need only restricted manipulation rights and obtain the scientific applications in the ready state to configure and submit them with minimal efforts;
(2) ``power users'' (``application developers'') --- the actual designers of the scientific processes and related workflows for simulations, usually they need to design and create the application (workflow) for comfortable work of the end-users.

\section{Use Cases of Practical Applications}
Several practical applications in materials science are presented below, where the different workflows were designed by WS-PGRADE workflow manager and used for computations in DCI by gUSE technology.

\subsection{Thermal Properties of Boron Nitride Nanotubes}
\begin{figure}[!t]
  \begin{center}
    \includegraphics[width=0.4\textwidth]{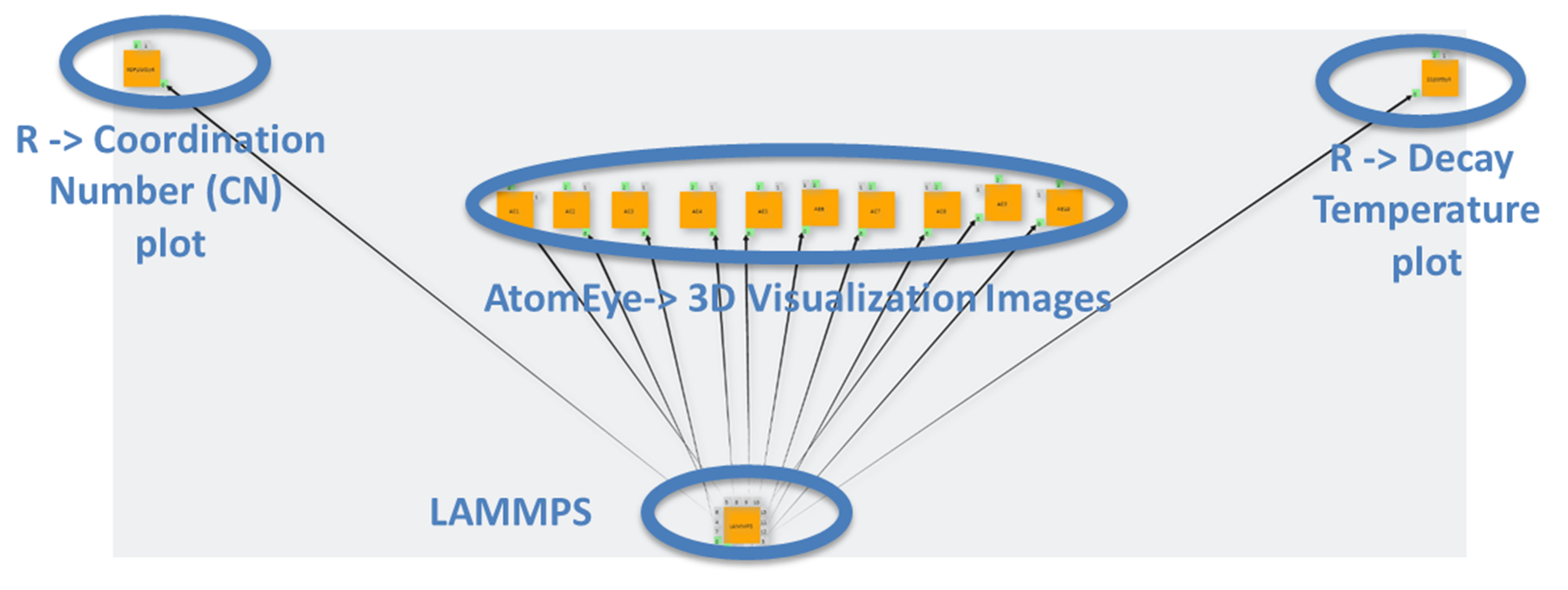} a)
    \includegraphics[height=0.2\textwidth]{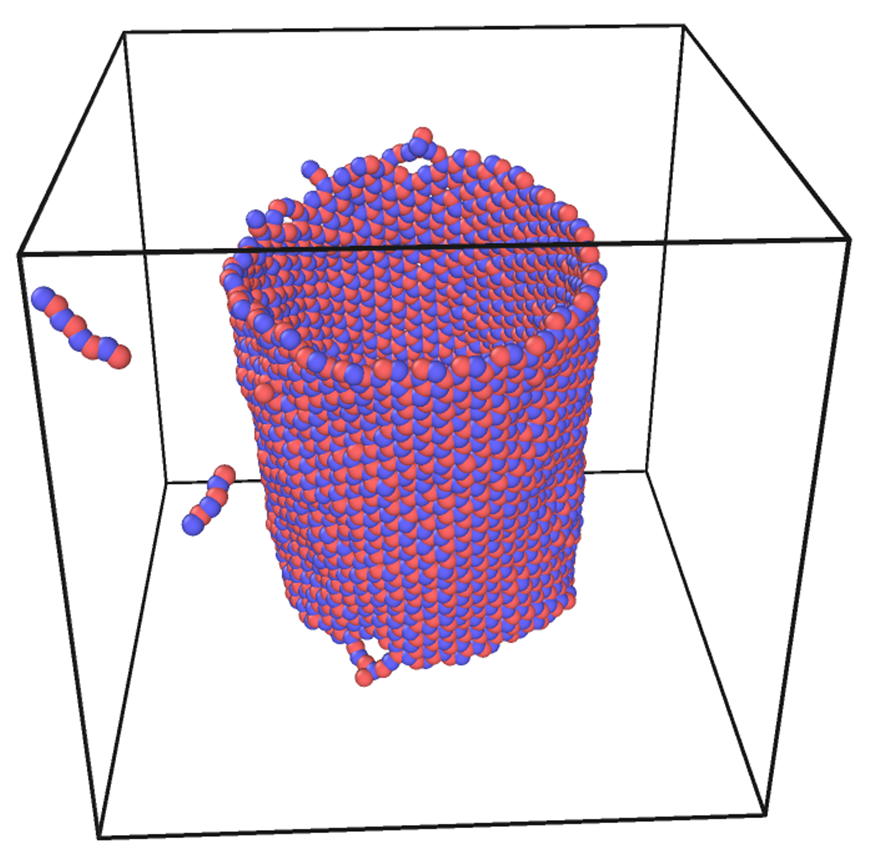} b)
    \includegraphics[height=0.3\textwidth]{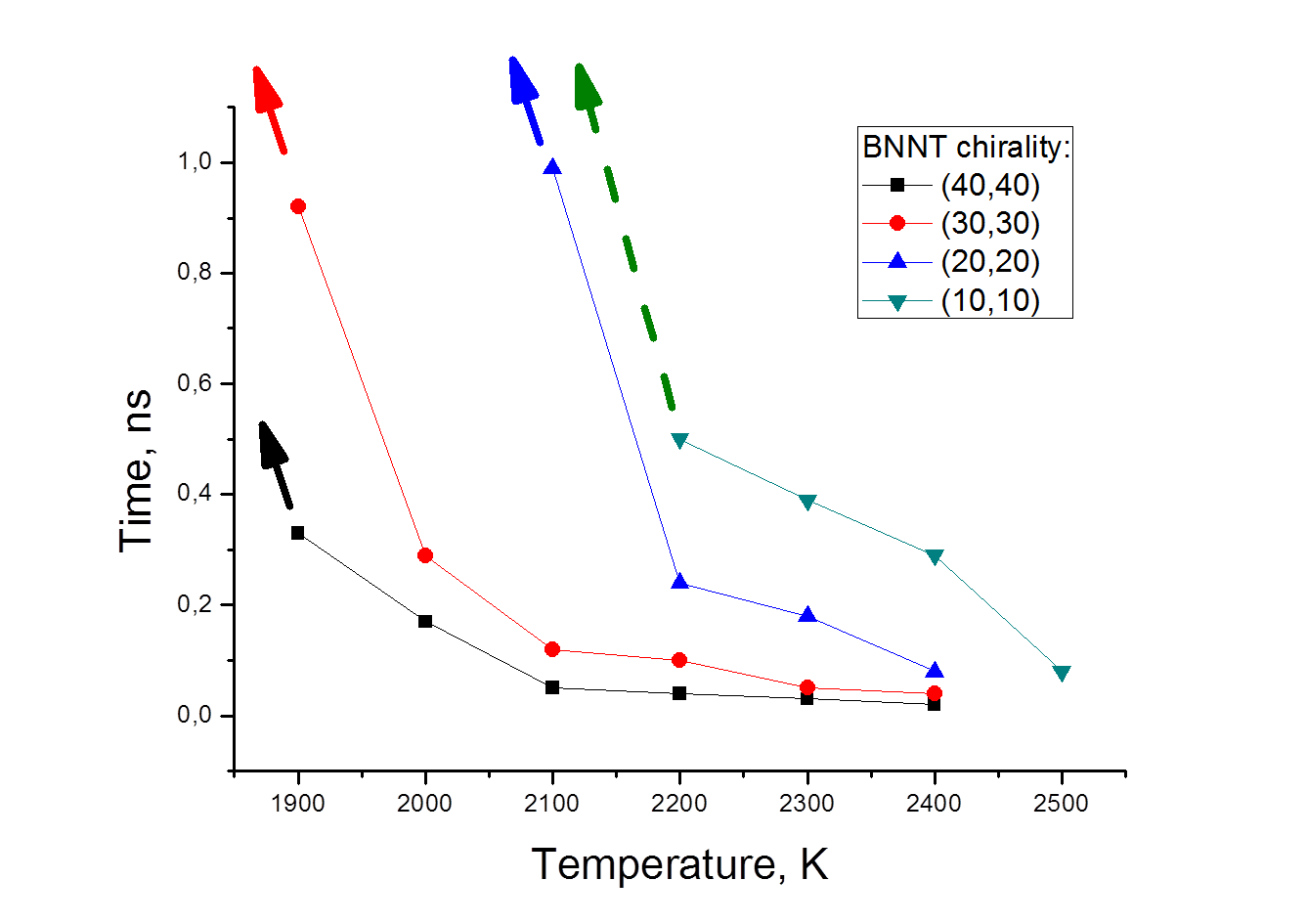} c)
    \caption{Use case for MD simulations of thermal properties of BNNTs: (a) the part of the general workflow (Fig.~\ref{fig:AL_WF_scheme}a), which was actually used here; (b) visualization by AtomEye (collapse of BNNT, nitrogen atoms --- red color, boron atoms --- blue color), (c) data post-processing and plot output by R-package (dynamics of the collapse for BNNT characterized by atom coordination numbers).}
    \label{fig:BNNT}
  \end{center}
\end{figure}

The scientific aim of this use case (Fig.~\ref{fig:AL_WF_scheme}) consists in MD simulations of boron nitride nanotubes (BNNT) with exceptional physical properties, which are a prerequisite for their wide practical applications in the future. The current user community in the field of nanotechnologies and materials science includes 2 user groups (4 end users) from IMP and IPMS (Frantsevich Institute for Problems in Materials Science, Kiev, Ukraine). MD simulation workflow (Fig.~\ref{fig:BNNT}a) was used for investigation of thermal stability of BNNT with different chirality, size and perfection of structure (Fig.~\ref{fig:BNNT}b). The actually used workflow (Fig.~\ref{fig:BNNT}a) was the part of the general workflow (Fig.~\ref{fig:AL_WF_scheme}a). It included the following subset workflows: 1) LAMMPS + R-package --- for MD simulations and plotting some post-processed values; 2) LAMMPS + R-package + AtomEye --- for the same operations + visualization of intermediate states of atoms.

From the physical point of view this work allowed us to derive some conclusions as to dynamics (Fig.~\ref{fig:BNNT}c) and temperature dependence of the BNNT collapse for various physical parameters (chiralities, sizes and perfection of structure) and the simulation parameters (type of potential, boundary conditions, thermal conditions, etc.). The obtained results extend the known experimental data on the thermal stability of BNNTs and described elsewhere \cite{nansys2013:sartinska}.

\subsection{Elastic Properties of Graphene}
\begin{figure}[!t]
  \begin{center}
    \includegraphics[width=0.4\textwidth]{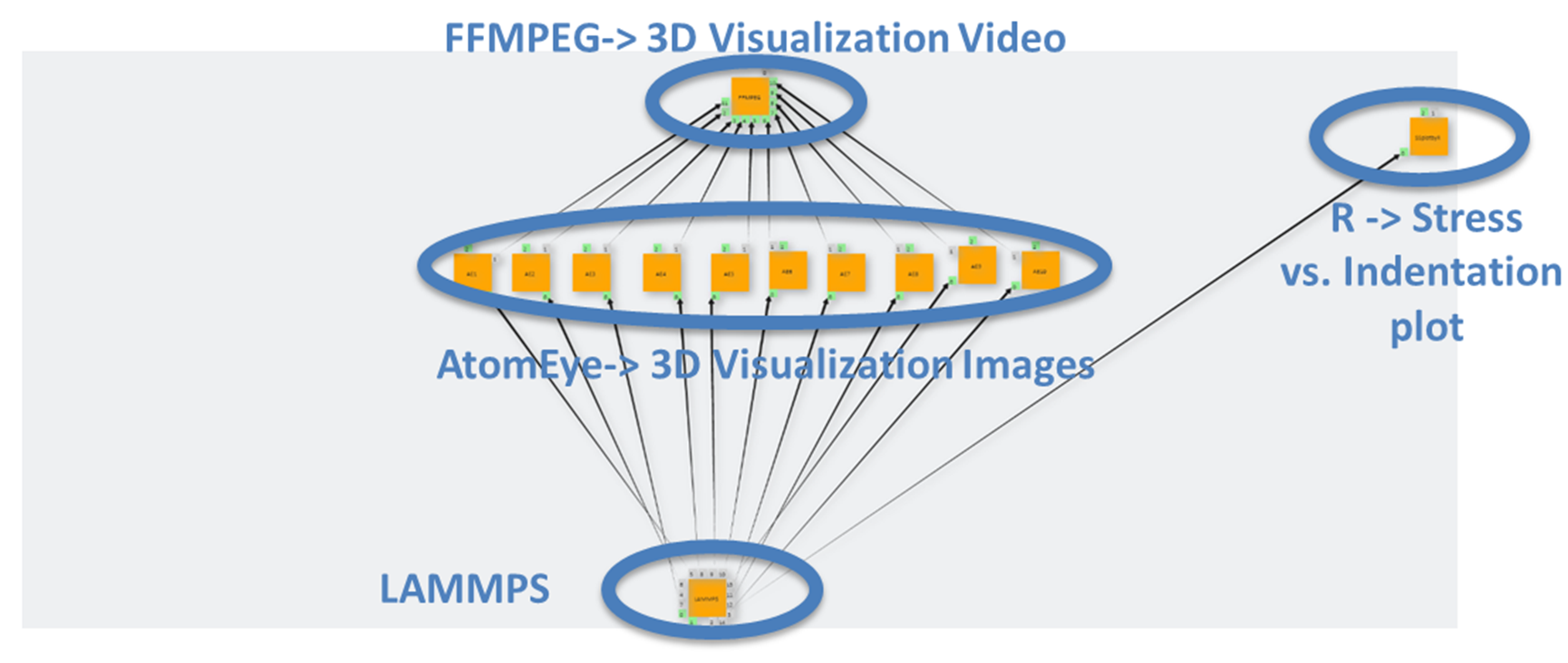} a)
    \includegraphics[height=0.2\textwidth]{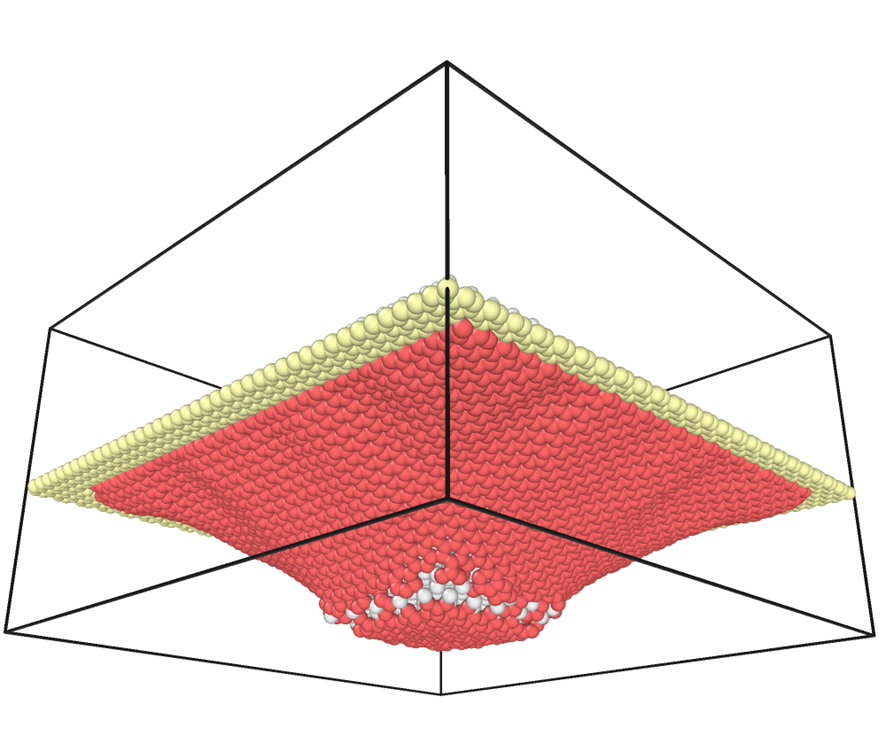} b)
    \includegraphics[height=0.3\textwidth]{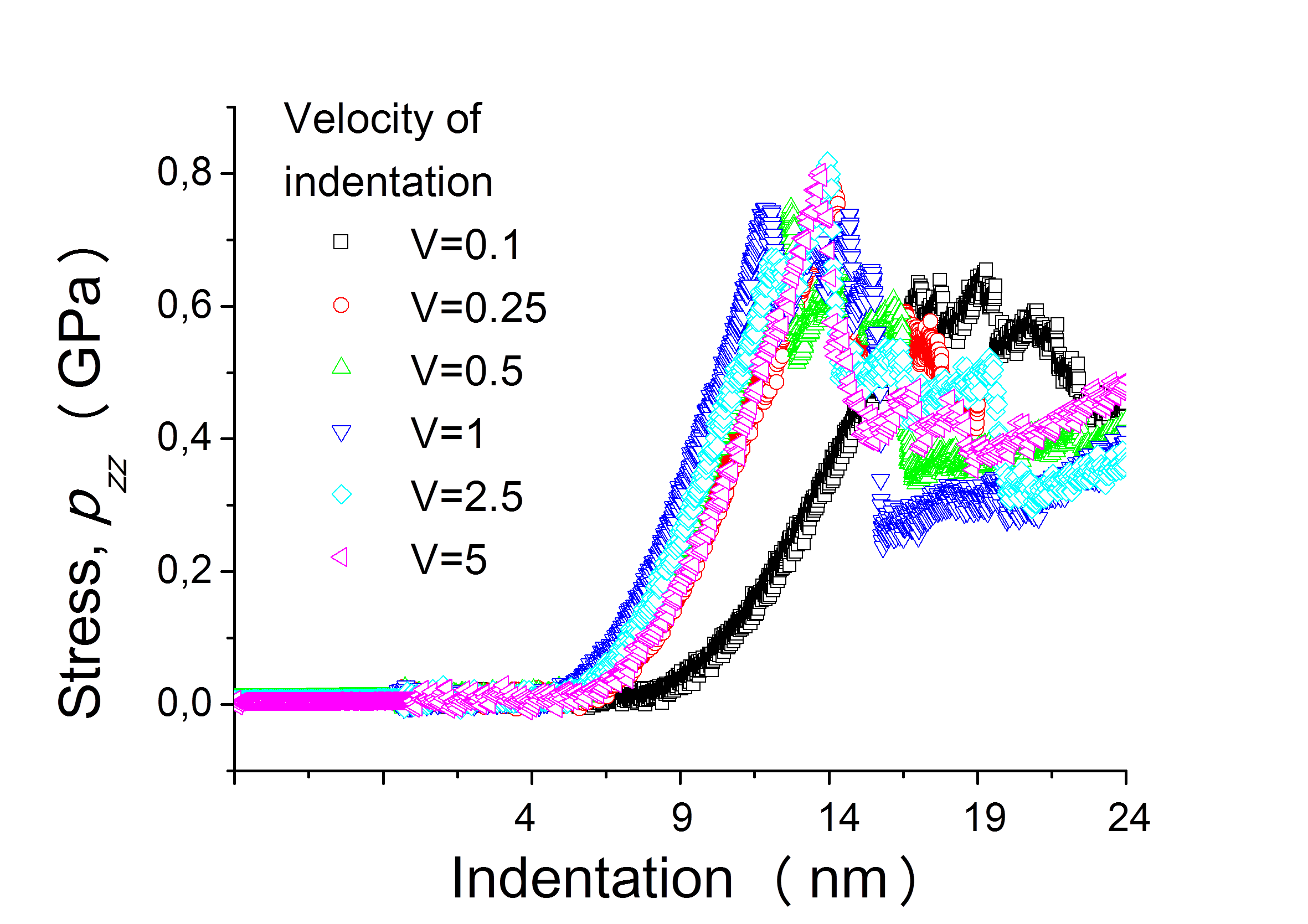} c)
    \caption{Use case for MD simulations of elastic properties of graphene: (a) the part of the general workflow (Fig.~\ref{fig:AL_WF_scheme}a), which was actually used here; (b) visualization by AtomEye (nanoindentation of the graphene monolayer, mobile carbon atoms --- red color, fixed carbon atoms --- yellow color, nanoindentor atoms --- gray color), (c) data post-processing and plot output by R-package (for different speeds of the nanoindentor).}
    \label{fig:nanoindenation}
  \end{center}
\end{figure}

The scientific aim of this use case (Fig.~\ref{fig:nanoindenation}) was MD simulation of nanoindentation of monolayer graphene membrane in an atomic force microscope. Its user community in the field of nanotechnologies includes 2 user groups (5 end users) from IMP and SPM/RS-Centre (Centre of scanning probe microscopy and resonance spectroscopy, Kyiv, Ukraine). MD simulation workflow (Fig.~\ref{fig:nanoindenation}a) was used for investigation of nanoindentation processes with different velocities, size of membranes and perfection of structure (Fig.~\ref{fig:nanoindenation}b). The actually used workflow (Fig.~\ref{fig:nanoindenation}a) was the part of the general workflow (Fig.~\ref{fig:AL_WF_scheme}a). It included the following subset workflows: 1) LAMMPS + R-package --- for MD simulations and plotting some post-processed values; 2) LAMMPS + R-package + AtomEye + FFMPEG --- for the same operations + visualization of intermediate states of atoms, and final video visualization.

From the physical point of view it was shown that graphene membrane deforms and collapses within a wide range of physical parameters (strain, applied force, indentor shape, indentor velocity, etc.) (Fig.~\ref{fig:nanoindenation}c). These results confirm and enrich the known experimental data of nanoindentation and described elsewhere \cite{nansys2013:tatarenko}.

\subsection{Mechanical Properties of Metal Nanocrystals}
\begin{figure}[!t]
  \begin{center}
    \includegraphics[width=0.4\textwidth]{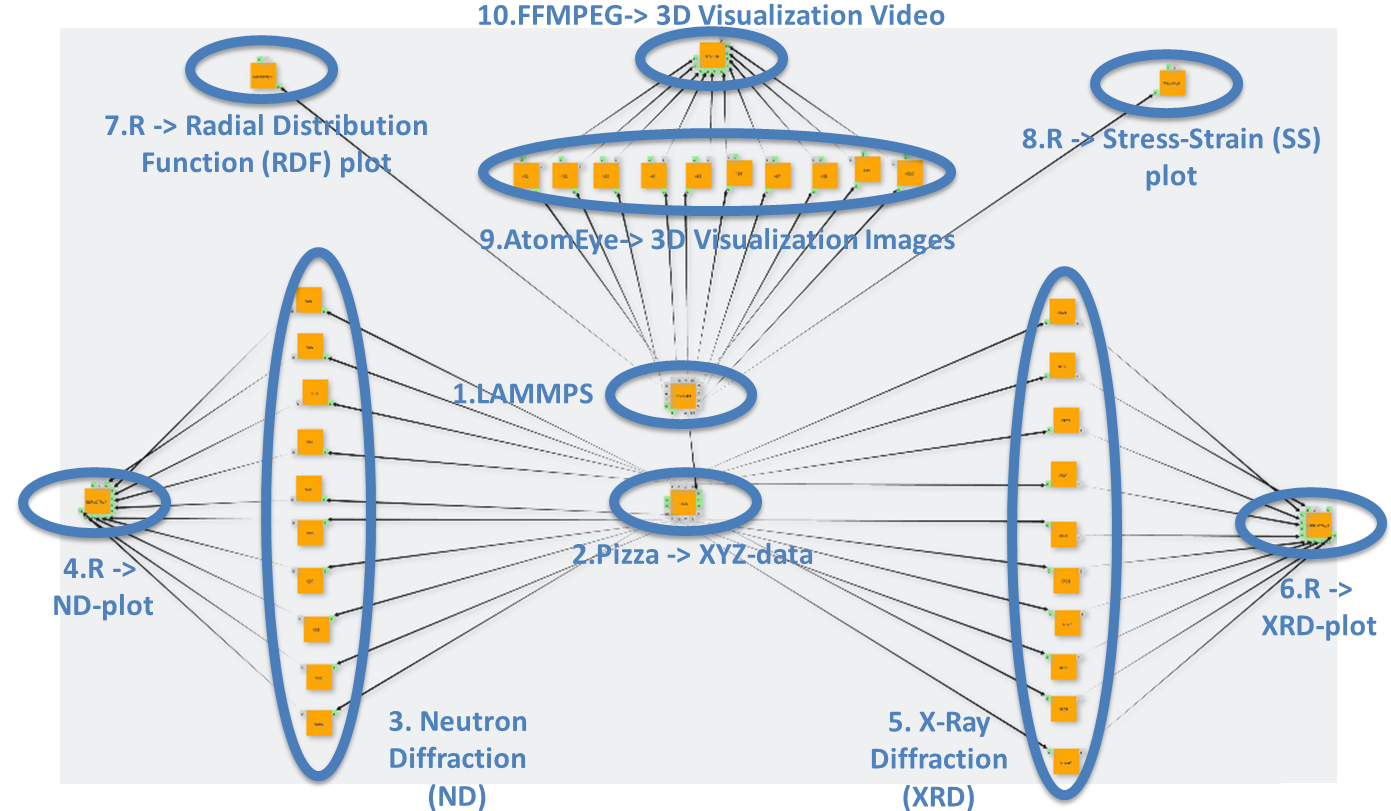}  a)
    \includegraphics[height=0.2\textwidth]{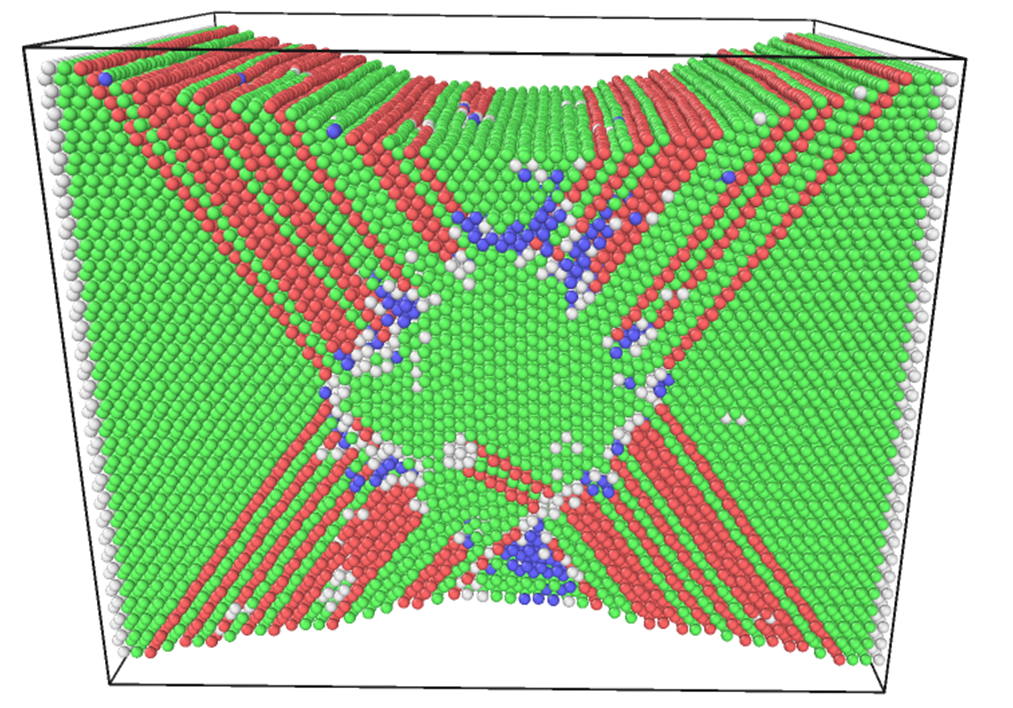} b)
    \includegraphics[height=0.3\textwidth]{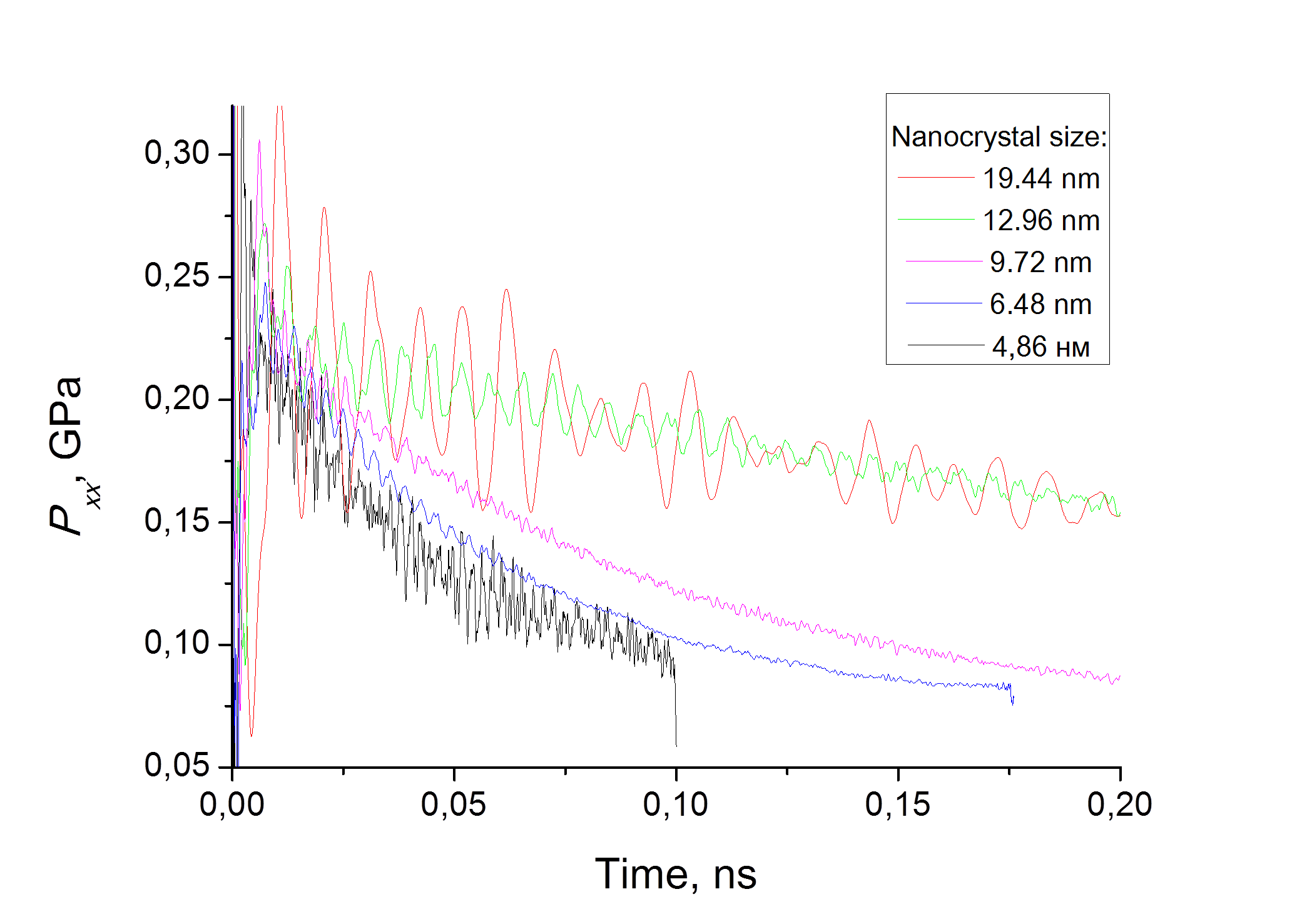} c)
    \includegraphics[height=0.3\textwidth]{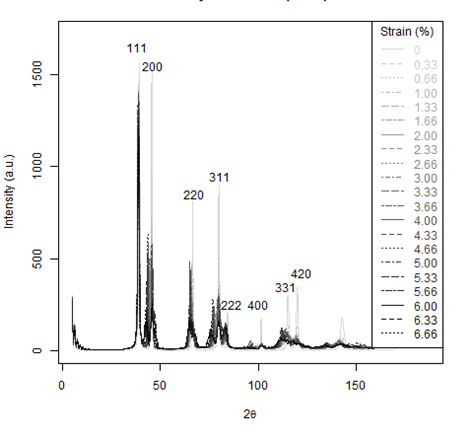} d)
    \caption{Use case for MD simulations of mechanical properties of metal nanocrystals: (a) the part of the general workflow (Fig.~\ref{fig:AL_WF_scheme}a), which was actually used here; (b) visualization by AtomEye (defect substructure evolution in Al nanocrystal, after 15 ps, atoms in ideal lattice --- green color, stacking faults --- red color), (c) data post-processing and plot output by R-package (oscillation of the internal stress $P_{xx}$ for different nanocrystal sizes), (d) data post-processing and plot output by R-package (X-ray diffraction analysis).}
    \label{fig:al_oscillations}
  \end{center}
\end{figure}

The scientific aim of this use case (Fig.~\ref{fig:al_oscillations}) was MD simulation of relaxation behavior of nanocrystals after uniaxial tension (Fig.~\ref{fig:al_oscillations}b) \cite{gordienko2011,dubna2012}. Its user community in the field of nanotechnologies includes 2 user groups (3 end users) from IMP and KNU (Kyiv National University, Kyiv, Ukraine). MD simulation workflow (Fig.~\ref{fig:al_oscillations}a) was used for investigation of relaxation behavior of nanocrystals for different crystals (Al, Cu, and Si), physical conditions (tensile rate, size and orientation of the nanocrystals) and methodological parameters (potential type, boundary conditions, and others). The actually used workflow (Fig.~\ref{fig:al_oscillations}a) includes the biggest parts of the general workflow (Fig.~\ref{fig:AL_WF_scheme}a). The simulations included the following subset workflows: 1) LAMMPS + R-package --- for MD simulations and plotting some post-processed values, like stress-strain dependencies (Fig.~\ref{fig:al_oscillations}c); 2) LAMMPS + R-package + AtomEye + FFMPEG --- for the same operations + visualization of intermediate states of atoms and final video visualization; 2) LAMMPS + R-package + AtomEye + FFMPEG + debyer --- for the same operations + processing data for X-ray and neutron diffraction and their plotting, like XRD-plots (Fig.~\ref{fig:al_oscillations}d).

From the physical point of view it was shown that the oscillatory relaxation of internal stresses is associated with the periodic rearrangement of the metastable defect substructure (Fig.~\ref{fig:al_oscillations}b). These results confirm and illustrate the known experimental data and described elsewhere \cite{nansys2013:steblenko}.

\section{Discussion}
The use cases presented here demonstrate that the general purpose and application specific workflows can be created in LEGO-style way by combining separate components-bricks, which corresponds to some real world operations, physical methods, and experimental devices even (see Table~\ref{table:VirtualLab}). Actually, such workflows can be easily designed and used in practice for MD simulations of complex natural processes in materials science, physics, and nanotechnologies. Each simulation workflow can be considered as a virtual counterpart (virtual lab) of the real experimental process, and the SG itself can be used as a hub of such virtual experimental labs in materials science.

\begin{table}[!h]
\caption{Results obtained, virtual lab (set of components used), and corresponding real lab}
\label{table:VirtualLab}
\centering
\begin{tabular}{|c|c|c|c|}
\hline
\textbf{Simulation result} & \textbf{Virtual lab,} & \textbf{Real lab,} \\
\textbf{(method needed)} &  \textbf{i.e. set of components} & \textbf{device}\\
\hline
Atomic structure & LAMMPS \cite{lammps1995} & JEOL R005\\
in bulk (TEM)       &  + AtomEye \cite{AtomEye2003}  & microscope \cite{jeol2010}\\
\hline
Stress-strain plot & LAMMPS \cite{lammps1995} & H-P testing machine, \\
(AFM)			 & + R \cite{R2008}  &  JEOL AFM \\
\hline
CN, RDF plots (neutron,  & LAMMPS \cite{lammps1995} &  X-ray\\
X-ray scattering)                 & + R \cite{R2008}    & diffractometer\\
\hline
X-ray diffraction  & LAMMPS \cite{lammps1995}    & SOLEIL \cite{soleil2005}\\
spectrum (XRD)  &  + debyer + R \cite{R2008}          & synchrotron facility \\
 \hline
Neutron diffraction & LAMMPS \cite{lammps1995} & Swiss Light Source \cite{sls2005}, \\
spectrum (ND)         & + debyer + R \cite{R2008}         & Large Hadron Collider \\
\hline
\end{tabular}
\end{table}

Some conclusions were made as to the possible future improvements of the workflow and resource management. The main problem was concerned with different demands for resources for various job parameters. For example, double increase of model size leads to nearly proportional double increase of the required memory and runtime. Sometimes this was in conflict with the available policies, like queueing rules, when the job runtime should be lower than the queue walltime. The typical example for the use case A (Fig.~\ref{fig:BNNT}) with the physical simulation time of 1~ns is shown in Table~\ref{table:Queues} for one of the used PBS-clusters, namely NTU HPCC (http://hpcc.kpi.ua) of NGI-UA. In this connection, the GUI-feature, which was pre-designed in SG for selection of a queue for PBS-clusters, was very useful, because it allowed to construct workflows optimized for runtimes lower that the prescribed queue walltime (or when the walltime is a multiple of the job runtime). This allowed to increase the overall efficiency, but the more progress could be reached by additional measures in SG, for example, by optimal job distribution over various queues (which should be automatic, and not manual, like here), and by polling all available queues among all available clusters with selection of the pool with the empty or more efficient resources.

\begin{table}[!h]
\caption{The queue rules and jobs in Use Case A (Fig.~\ref{fig:BNNT})}
\label{table:Queues}
\centering
\begin{tabular}{|c|c|c|c|}
\hline
\textbf{Queue name} & \textbf{Walltime} & \textbf{Cores/User} & \textbf{Max BNNT (atoms)} \\
\hline
ku-small  & 90 min.  & 32  & $<10\times10$ $(<840)$ \\
\hline
ku-single & 8 days   & 4   & $>40\times40$ $(>3360)$ \\
\hline
ku-normal & 180 min. & 32  & $<30\times30$ $(<2520)$ \\
\hline
kh-large  & 120 min. & 128 & $<20\times20$ $(<1680)$ \\
\hline
\end{tabular}
\end{table}

Other means for increase of SG-efficiency, like dual-layer hardware and software management in cluster systems \cite{stirenko2013}, will be also of interest and great value and should be taken into account. Additional important aspect is related with the failed jobs, which should be re-run. The problem is that for the big number of jobs (each for different set of parameters) it was not very easy to filter the faulty jobs and re-run them again. At this stage, it was performed manually by means of job management system, but it would be more efficient, if this will done automatically by SG after obtaining the corresponding report.

\section{Conclusions}
Science Gateway Portal (like ``IMP Science Gateway Portal'') for MD simulations of complex behavior of various nanostructures can be relatively easily created on the basis of WS-PGRADE technology for workflow management. These resourceful MD simulations can be effectively carried out in the heterogeneous DCI by means of gUSE technology for parallel and distributed computing in various heterogeneous DCIs. Several typical scientific applications were considered as use cases of their porting and practical using. Several advantages should be emphasized: (1) the modular-based (``LEGO-style'') principle of workflow creation and management; (2) easy access to heterogeneous DCI and software; (3) division of user roles (administrators as operators of portals; power users as principal scientists; end users as scientists, students); (4) short learning curve for usual scientists without extensive knowledge in computer science. Some ways for further improvement were briefly outlined in relation to the better performance and efficiency. Finally, the ``science gateway'' approach -- workflow manager (like WS-PGRADE) + DCI resources manager (like gUSE) at the premises of the portal (like ``IMP Science Gateway Portal'') -- is very promising in the context of its practical MD applications in materials science, physics, chemistry, biology, and nanotechnologies.

\section*{Acknowledgments}
The work presented here was partially funded by EU FP7 SCI-BUS (SCIentific gateway Based User Support) project, No. RI-283481, and partially supported in the framework of the research theme "Introduction and Use of Grid Technology in Scientific Research of IMP NASU" under the State Targeted Scientific and Technical Program to Implement Grid Technology in 2009-2013.



%

\end{document}